\documentclass[%
aip,
 amsmath,amssymb,
 reprint,
 nofootinbib,
 twocolumn
twocolumn]{revtex4-1}
\usepackage{amsmath}
\usepackage{mathtools}
\usepackage{amssymb}
\usepackage{physics}
\usepackage{color}
\usepackage{graphicx}
\usepackage{tabularx}
\usepackage{dcolumn} 
\usepackage{bm}      
\usepackage{array}   
\usepackage{amsfonts}
\usepackage[bookmarks=false,colorlinks,citecolor=red]{hyperref}

\usepackage[utf8]{inputenc}
\usepackage[T1]{fontenc}
\usepackage{mathptmx}
\usepackage{xcolor}

\usepackage{subfiles}

\include{MyCommand}

\DeclareUnicodeCharacter{2212}{-}

\newcommand{\al}{\alpha}

\makeatletter
\newcommand*{\addFileDependency}[1]{
  \typeout{(#1)}
  \@addtofilelist{#1}
  \IfFileExists{#1}{}{\typeout{No file #1.}}
}
\makeatother

\include{MyCommand}

\usepackage{hyperref}
\hypersetup{
    colorlinks=true,
    linkcolor=blue,
    filecolor=magenta,      
    urlcolor=cyan,
    pdftitle={Overleaf Example},
    pdfpagemode=FullScreen,
    }

\begin{document} 


\title{VMC Optimization of Ultra-Compact, Explicitly-Correlated Wave Functions of the Li Isoelectronic Sequence in Its Lowest 1s2s2p Quartet State 
} 




\author{D~.~J.~ Nader}
 \email{daniel_nader@brown.edu} 
 \affiliation{Department of Chemistry, Brown University, Providence, Rhode Island 02912, United States}
  \author{B.~M.~ Rubenstein}
  \email{brenda_rubenstein@brown.edu}
   \affiliation{Department of Chemistry, Brown University, Providence, Rhode Island 02912, United States}
 
\begin{abstract}
A compact yet accurate approach for representing the wave functions of members of the He and Li isoelectronic series is using explicitly correlated wave functions. These wave functions, however, often have nonlinear forms, which make them challenging to optimize. In this work, we show how Variational Monte Carlo (VMC) can efficiently optimize explicitly correlated wave functions that accurately describe the quartet 1$s$2$s$2$p$ state of the Li isoelectronic sequence with ten or fewer parameters. We find that our compact wave functions correctly describe cusp conditions and reproduce at least 99.9\% percent of the exact energy. 
\end{abstract}


\maketitle

\section{\label{sec:intro}Introduction}

The isoelectronic sequence of lithium, consisting of atoms with three electrons and a nucleus, is the simplest sequence of atomic systems that have both open and closed shell ground states, and thus serve as prototypes for the more complex ground states of heavier alkali-metal atoms and alkaline-earth cations.\cite{THAKKAR200295} The non-relativistic wave functions of atoms and ions with three electrons such as those in this sequence are also of great interest to those developing highly accurate evaluations of relativistic and QED corrections.\cite{Ruiz2013,Yerokhin2021} The Li isoelectronic sequence has therefore become a testbed for quantum chemical methods, much like He-like atoms were for many decades after the advent of quantum mechanics. 

One particularly interesting state exhibited by members of this series is the 1$s$2$s$2$p$ state, which is the lowest electronic state of the quartet manifold with total spin $S=3/2$. Since optical quartet-to-doublet transitions are difficult to
observe experimentally, the experimental uncertainty of this state is too large for it to be used in Grotrain diagrams.\cite{Sven1983} This state of affairs has motivated researchers to develop more accurate electronic structure techniques and other means for determining the term values of this quartet state during the end of the last century.\cite{Hsu1991tt} 
The confirmation that a positron can be attached to this state to form $e^+{\rm Li}$ renewed interest in its electronic structure some years later.\cite{Bubin2013,Yamashita,Bressanini} 
Additionally, the 1$s$2$s$2$p$ state is an eigenfunction of the angular
momentum operators $\hat{L}^2$ and $\hat{L}_z$ with the eigenvalues $L(L + 1) = 2$ and $M_L = 0$. Its ground state wave function is therefore not only a function of interelectronic distances (which would make its electronic structure problem into a six-dimensional problem), but is also dependent upon angular contributions that make it into a nine-dimensional problem. 

Traditionally, the ground state of such species would be described using a single determinant with optimized orbitals or a multideterminant expansion with linear coefficients. Such linear expansions are often preferred because they lead to 
highly accurate energies, can be solved through standard linear algebra techniques, and automatically provide one with a physical picture of which electronic excitations contribute most to a state based upon the coefficients obtained.  Previous works using linear expansions have provided benchmark energies for the $1s2s2p$ state: Conventional Configuration Interaction (CI) \cite{Bunge,Lunell1977,Hsu1991tt,Hsu1994} with 2519 determinants and the Hylleraas method \cite{Larsson1982,BARROIS1997531} with 1372 Hylleraas functions yielded energies with $\mu$Ha accurate, as summarized in Ref. \onlinecite{KING199957}.  However, given the number of determinants that were needed to accurately describe this state, alternative approaches become attractive.

One set of alternative approaches are so-called explicitly-correlated approaches that attempt to find the ``most compact" representations of wave functions, sometimes at the expense of overall accuracy. \cite{Bressanini_2008,David2006,Harris2005,THAKKAR200295} \footnote{Note that these explicitly-correlated methods differ from the explicitly-correlated F12 and R12 methods of recent note in the literature.\cite{kong2012explicitly}} Compact, yet accurate wave functions are particularly valuable  if they accurately reproduce cusp conditions. From a practical point of view, explicitly-correlated compact wave functions are often used in the study of collisions \cite{Jones2003,Ancarani_2008,Ciappina,Kircher2022,Madison2003,Bahati_2005,Defrance_2000} or to ease the computation of the matrix elements of the numerous singular operators representing relativistic and QED corrections. \cite{Yerokhin2021} One of the best approaches for designing compact wave functions is to include explicit correlation and use nonlinear variational parameters.\cite{Kong2012} However, in contrast with conventional linear expansions in terms of determinants, the optimization of such wave functions is challenging since the optimal parameters can not be found by solving the secular Schr\"odinger equation. The potentially large number of nonlinear parameters additionally presents a steep challenge for standard minimization algorithms. 

Given this backdrop, a potentially promising technique for optimizing these challenging wave functions is the Variational Monte Carlo (VMC) method. In the VMC method, the parameters within a given wave function ansatz are optimized to minimize the energy by iteratively using Monte Carlo sampling to evaluate the energy for a given set of parameters and then finding an improved set of parameters that further minimize the energy.\cite{TOULOUSE2016285,Rubenstein2017} While the energies given the parameters could be evaluated using traditional grid-based integration methods, this becomes more computationally costly than random sampling when eight or more dimensions are involved. Monte Carlo methods thus become the methods of choice for high dimensional search spaces.\cite{TOULOUSE2016285} Although VMC is most popularly used to optimize the Jastrow parameters within Slater-Jastrow wave functions,\cite{Needs_2010} it is equally applicable to wave functions with other forms, including explicitly-correlated wave functions. 

In this manuscript, we thus employ VMC to optimize explicitly correlated wave functions containing 7 and 10 parameters to describe the  lowest quartet state of the Li isoelectronic sequence. We show that VMC is able to rapidly find and converge parameters to a set that minimizes the energy. We find that the optimal parameters in general fit to Pad\'e functions of the nuclear charge within the error bars.  Our VMC-optimized wave functions yield energies which reproduce at least 99.9\% of the most accurate Hylleraas results. Our work therefore demonstrates the potential that VMC methods have for optimizing explicitly-correlated wave functions of difficult to describe, higher angular momentum states.

\section{\label{wf} Wave Function Describing the Li Isoelectronic Series 1$s$2$s$2$p$ Quartet State}

\begin{figure*}
    \includegraphics[width=10cm, height=6cm,angle=0]{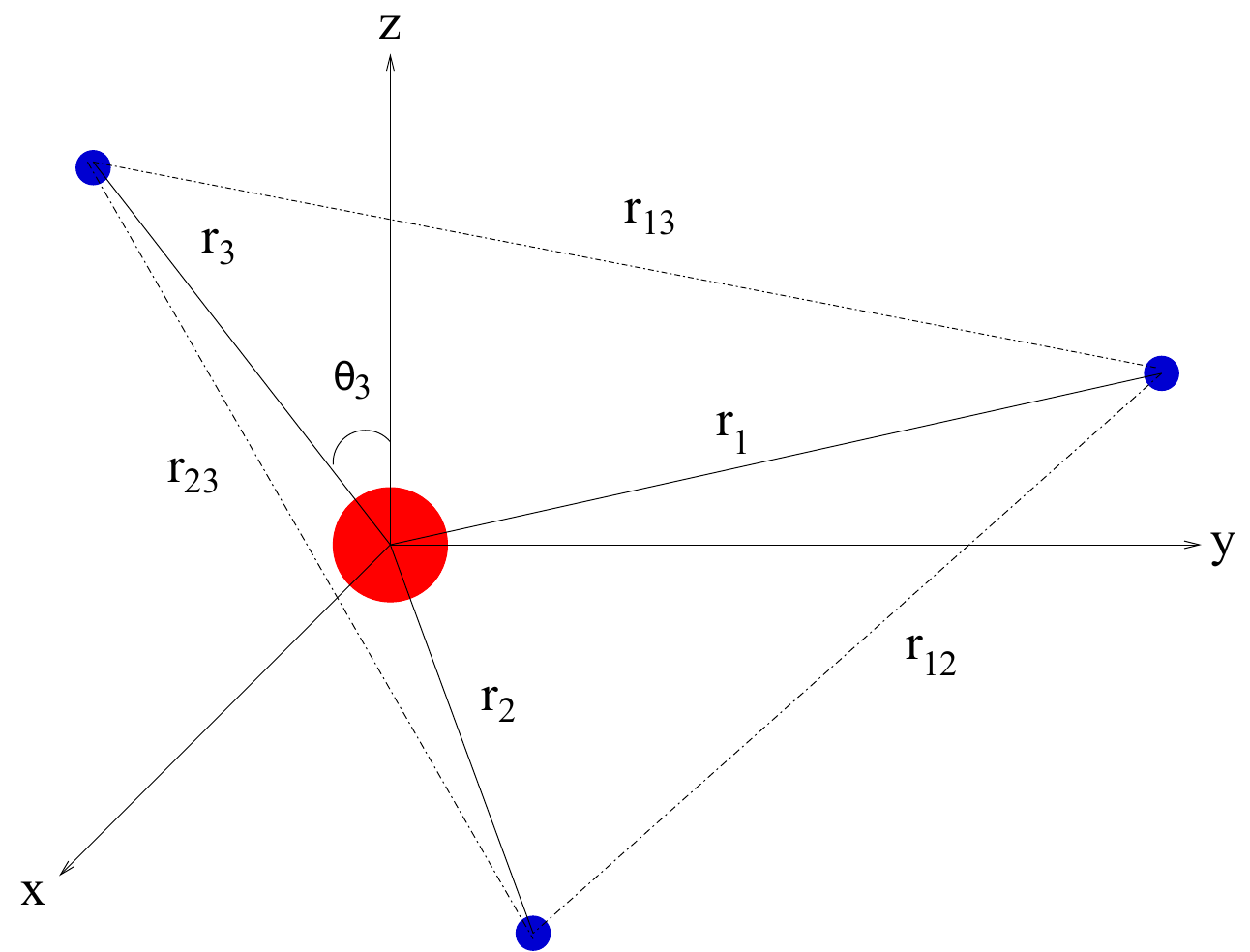} 
  \caption{\label{Geometry}
    Geometry of a representative member of the lithium isoelectronic sequence consisting of three electrons (blue) and a nucleus (red). Key distances and bond angles, some of which are used in our explicitly correlated wave functions, are denoted in black. %
    }
\end{figure*}

Members of the lithium isoelectronic sequence contain a total of four charged subatomic particles - three electrons and one nucleus - interacting via the Coulomb potential. Since the motion of the nucleus is much slower than that of the electrons, in the limit of infinite mass, it can be assumed to be a fixed, positively-charged center, effectively reducing the problem from twelve to nine degrees of freedom. We illustrate such a system, its parameters, and the notation we later employ for those parameters in Figure \ref{Geometry}.

Based on the conventional atomic shell model, the ground state with total spin $S=1/2$ possesses two electrons that occupy the same orbital and a third that occupies a different, higher-energy orbital. However, for quartet states with total spin $S=3/2$, all of the electrons lie in different orbitals. This state is harder to describe because of its higher-spin and open shell character.
 
To design an explicitly-correlated wave function that can describe this state, we first consider the physics of these systems when the positive nuclear charge, $Z$, goes to infinity $Z\to\infty$. In this limit, nuclear-electron interactions dominate and the electron-electron interactions can be neglected, reducing the problem to that of three hydrogen electrons in three different orbitals. In this limit, there exists an exact solution to the Schrödinger equation in the form of an anti-symmetrized product of three Coulomb orbitals 

\begin{eqnarray}
\label{wfmotivacion}
 1s_1&\sim&e^{-\alpha_1 r_1}\nonumber \\
 2s_2 &\sim&(1+ar_2) e^{-\alpha_2 r_2} \nonumber \\
  2p_3&\sim&  r_3\cos\theta_3 e^{-\alpha_3 r_3} \nonumber\\
\psi_0&\sim&1s_12s_22p_3\nonumber \\
\psi_0&\sim&(1+ar_2)(r_3\cos\theta_3)\times \nonumber\\
& & e^{-\alpha_1  r_1 - \alpha_2  r_2 - \alpha_3  r_3} \,,
\end{eqnarray}
where $\alpha_1=Z$, $a=-Z/2$ and $\alpha_2=\alpha_3=Z/2$.\cite{Stillinger} $r_i$ and $r_{ij}$ are the relative distances defined in Figure \ref{Geometry}.

Motivated by Equation (\ref{wfmotivacion}), we follow a recipe for designing compact wave functions described in our previous work.\cite{Turbinercompact1,Turbinercompact2,TurbinerCompact3} The total wave function, $\Psi$, is the antisymetrized product of the spatial wave function, $\psi$, and the spin function, $\chi$, $\Psi=\mathcal{A}(\psi\chi)$. The antisymmetrization operator is defined as
$$\mathcal{A}=1-\hat{P}_{12}-\hat{P}_{13}-\hat{P}_{23}+\hat{P}_{231}+\hat{P}_{312}\,,$$
where $\hat{P}_{ij}$ permutes the electrons $i\leftrightarrow j$ and $\hat{P}_{ijk}$ permutes the $(ijk)$ indices. Since the spin function for the quartet states is totally symmetric, $\chi=\alpha\alpha\alpha$, where $\alpha$ denotes 
the spin-up function, the anti-symmetrization ends up being only performed on the spatial term $\psi$. 
The spatial wave function we propose has the following form

\begin{eqnarray}
\label{wfa}
\psi&=&(1+ar_2)\left(r_3\cos\theta_3\right)\times \nonumber\\
& &
e^{-\alpha_1 r_1-\alpha_2 r_2-\alpha_3 r_3+\alpha_{12} r_{12}+\alpha_{13} r_{13}+\alpha_{23} r_{23}}\,,
\end{eqnarray}
where $\alpha_i$ with $i=1,2,3$, $\alpha_{ij}$ with $j > i=1,2,3$, and $a$ are free variational parameters. 
The nonlinear variational parameters $\alpha_i$ and $\alpha_{ij}$ represent screening/antiscreening factors for the Coulomb charges in the nucleus-electron and electron–electron interactions,
respectively. The factor depending on $\cos\theta_3$ leads to the required
odd parity, while the total angular momentum $L=1$ guarantees orthogonality with the 1$s$2$s$3$s$ quartet state. We denote the wave function in Equation (\ref{wfa}), containing seven variational parameters, as Ansatz A. 

As a generalization, we also consider inserting the following rational expressions into the exponents of the orbitals
\begin{equation}
\label{sustitution}
\alpha_i r_i \to \alpha_i \hat{r}_i=\alpha_i r_i\frac{1+c_i r_i}{1+d_i r_i}\,. 
\end{equation}
These terms interpolate the effective Coulomb charges between their values at small and large distances. Calculations show that only some terms in the exponential lead to a significant difference, beyond statistical fluctuations, when replaced by Equation (\ref{sustitution}). This leads to Ansatz B, given by 
\begin{eqnarray}
\label{wfb}
\psi&=&(1+ar_2)\left(r_3\cos\theta_3\right)\times \nonumber \\
& &e^{-\alpha_1 r_1-\alpha_2 r_2-\alpha_3 \hat{r}_3+\alpha_{12} \hat{r}_{12}+\alpha_{13} r_{13}+\alpha_{23} r_{23}}\,,
\end{eqnarray}
which has 11 variational parameters. Our variational calculations show that $\alpha_3$ can be fixed to $Z/2$, with no impact on the variational energy, effectively reducing the number of variational parameters to 10. This can be understood because, upon introducing the substitution $\al_3r_3\to \al_3\hat{r}_3$, the screening charge of the nucleus as seen from the third electron at short distances is $Z$, leading $\alpha_3\to Z/2$  according to Equation (\ref{wfmotivacion}).

Given these numbers of parameters, in the following, we optimize the parameters for the 7-parameter Ansatz A and the 10-parameter Ansatz B using Variational Monte Carlo.

\section{\label{vmc} Variational Monte Carlo}

Variational Monte Carlo \cite{Rubenstein2017,Foulkes2001,TOULOUSE2016285} is based upon the variational principle, which states that the variational energy of a system is given by 
\begin{equation}
\label{Evar}    
E_v=\frac{\langle \Psi | \hat{H}|\Psi \rangle }{\langle \Psi|\Psi\rangle}= \frac{\sum_\sigma \int d{\bf R} \Psi^\dagger({\bf R},\sigma)\hat{H} \Psi({\bf R},\sigma)}{\sum_\sigma \int d{\bf R} \Psi^\dagger({\bf R},\sigma) \Psi({\bf R},\sigma) }\leq E_0\,,
\end{equation}
where ${\bf R}=\{{\bf r}_1,{\bf r}_2,{\bf r}_3\}$ and $\sigma={\sigma_1,\sigma_2,\sigma_3}$ denote the positions and spins of the three particles (here, electrons), respectively, and $E_0$ is the exact ground state energy. 
The variational energy, $E_v$ is estimated via the numerical evaluation of a  nine-dimensional integral over the nine particle positions which can be performed via Monte Carlo methods.

To make Equation (\ref{Evar}) more directly amenable to Monte Carlo sampling, it can be rewritten as
\begin{equation}
\label{EvarMC}    
E_v=\frac{\sum_\sigma \int d{\bf R} P({\bf R},\sigma) \frac{\hat{H} \Psi({\bf R},\sigma)}{\Psi({\bf R},\sigma)}}{\sum_\sigma \int d{\bf R} P({\bf R},\sigma) }\,,
\end{equation}
where $P({\bf R},\sigma)$ is a probability distribution of the form $P({\bf R},\sigma)=\Psi^\dagger({\bf R},\sigma)\Psi({\bf R},\sigma)$. The quantity $ \frac{\hat{H} \Psi({\bf R},\sigma)}{\Psi({\bf R},\sigma)}$ is known as the local energy, $E_L$. In the VMC method, $P$ is sampled to generate a set of $N$ configurations in $\{{\bf R}, \sigma \}$ space that are used to estimate the integral above. A common way to sample such configurations is provided by the Metropolis algorithm. The spin variable, $\sigma$, can be integrated out analytically, both in the numerator and denominator of Equation (\ref{EvarMC}). This is because, for quartet states, there is only one spin function corresponding to the total spin $S=3/2$ and projection $S_z$, on which the Hamiltonian does not act. The local energy is then evaluated based upon the configurations sampled and averaged to approximate the variational energy. 

Other expectation values can also be calculated within the VMC framework by replacing the local energy in Equation (\ref{EvarMC}) by the quantity of interest. In particular, we are interested in computing the following ratio between expectation values
\begin{equation}
\label{cusp}
C_{Ne}=\frac{\langle \delta({\bf r}_i) \frac{\partial}{\partial r_i}\rangle}{\langle \delta({\bf r}_i)\rangle}\,, 
\end{equation}
in order to estimate the cusp condition $C_{Ne}\approx Z$, as a measure of the quality of the wave function in the vicinity of the Coulomb singularities.

The accuracy of the variational energy and other quantities so obtained depends upon the wave function parameterization, which can be optimized as described below. For a more detailed review of the VMC method, we refer the reader to Refs. \onlinecite{TOULOUSE2016285,Pang2016,Foulkes2001,Hammond1994,Rubenstein2017}.

\section{Optimization of the Wave Function Ansatz}

In order to reach the lowest variational energy possible given the wave function ansatz, the wave function's parameters must be optimized. This can be achieved by performing gradient descent based upon the variational energies obtained using the VMC sampling described above. The derivative of the energy with respect to the variational parameters $p$ is given by\cite{Rubenstein2017}
\begin{eqnarray}
\label{derivatives}
\frac{\partial \langle \hat{H}\rangle }{\partial p}&=& \frac{\langle \Psi | \Psi\rangle\frac{\partial}{\partial p }\langle \Psi|\hat{H}|\Psi\rangle -\langle \Psi|\hat{H}|\Psi\rangle \frac{\partial}{\partial p }\langle \Psi | \Psi\rangle}{\langle \Psi | \Psi\rangle^2} \nonumber \\
&=&\frac{\langle \partial_p \Psi| \hat{H}|\Psi \rangle + h.c. }{\langle \Psi | \Psi\rangle}-\langle \hat{H}\rangle\frac{\langle \partial_p\Psi|\Psi\rangle + h.c.}{\langle \Psi | \Psi\rangle}
\end{eqnarray}
Within the framework of VMC, we can calculate the derivatives in Equation (\ref{derivatives}) using the equations
\begin{eqnarray}
\label{ExVMC}
\langle \hat{H}\rangle &=&  \int \Psi^*\Psi \frac{\hat{H}\Psi}{\Psi}d{\bf R}= \int 
P({\bf R}) E_L d{\bf R} \\ 
\langle \Psi | \partial_p\Psi\rangle &=& \int \Psi^*\Psi \frac{\partial_p\Psi}{\Psi}d{\bf R}= \int 
P({\bf R}) F_L^{p} d{\bf R} \\
\langle \Psi|\hat{H}| \partial_p\Psi\rangle &=&  \int \Psi^*\Psi \frac{\hat{H} \partial_p\Psi}{\Psi}d{\bf R}= \int 
P({\bf R}) G_L^{p} d{\bf R}\,,
\end{eqnarray}
and evaluate the related integrals in one single calculation based on the Metropolis algorithm. We update the variational parameters $p_{k+1}$ corresponding to the next iteration $k+1$ according to
\begin{equation}
p_{k+1}=p_k+\lambda_p\left( \frac{\partial \langle \hat{H}\rangle}{\partial p}\right)_k\,
\end{equation}
where $\lambda_p$ is an update rate which can be tuned for each variational parameter $p$ in order to accelerate convergence.

\section{Numerical Details}

To perform these VMC calculations, we  wrote a Fortran-based code 
which optimizes wave functions in the form of Ansatz A (Equation \ref{wfa}) and Ansatz B (Equation \ref{wfb}) for lithium in its 1$s$2$s$2$p$ quartet state. 
The optimization is carried out using the gradient descent method during which VMC is used to determine energies and gradients and stochastic gradient is iteratively used to update the parameters based upon these gradients. One iteration typically consists of 10$^{7}$ Metropolis steps. We view the wave function as optimized when the variational energy converges, i.e., does not further decrease, within statistical error bars, after 50 consecutive iterations. It is worthwhile to mention that the optimization can be started from different starting configurations of parameters to reduce the risk of finding a local minimum. 
The number of iterations needed to obtain energy convergence, i.e., when all variational parameters reach equilibrium and oscillate around their optimal values, highly depends on the descent parameter $\lambda$. Practical calculations show that some parameters reach equilibrium faster than others.
Thus, we used different descent parameters for the different wave function parameters: $\lambda_{\alpha_i}=2$ for $i=1,2,3$, $\lambda_{\alpha_{ij}}=0.2$ for $i<j$, and  $\lambda_{a}=10$. With these descent parameters, we observe convergence after $\sim50$ iterations (see Fig. \ref{optimization}, for instance) and compute expectation values after this convergence has been attained. We wrote a specialized code to estimate the cusp parameter Equation (\ref{cusp}) once parameters have been optimized and expectation values have been converged.  

\section{Results}

\subsection{Variationally Optimized Wave Functions and Energies}

\begin{figure*}
    \includegraphics[width=10cm, height=8cm,angle=0]{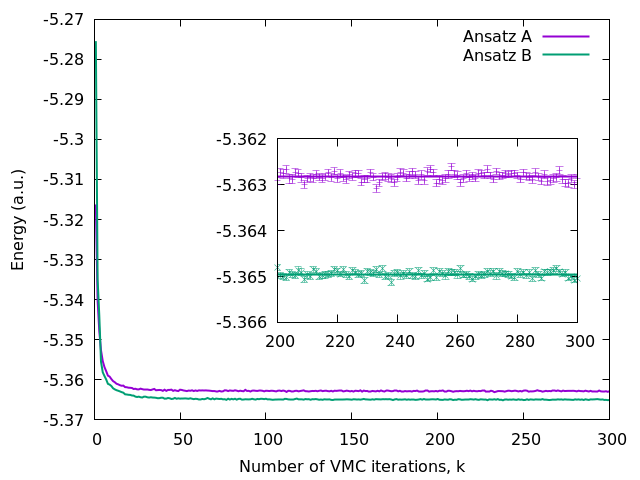} 
  \caption{\label{optimization}
    Energy of neutral Lithium in the $1s2s2p$ state as a function of the number of VMC iterations. Ansatz A energies and error bars are given in purple, while Ansatz B energies and error bars are given in green.  
    }
\end{figure*}

Using the VMC approach described above, we optimized the explicitly-correlated wave functions given by Ansatz A (Equation \ref{wfa}) and Ansatz B (Equation \ref{wfb}) for neutral lithium and its isoelectronic sequence with nuclear charges between $3-10$ in the 1$s$2$s$2$p$ state. A good initial set of variational parameters is that given by the analytical solution for the infinite nuclear charge (Equation \ref{wfmotivacion}) i.e.,  $\alpha_1=Z$, $a=-Z/2$, $\alpha_2=\alpha_3=Z/2$, and $\alpha_{ij}=0$.
For instance, we show in Figure \ref{optimization} the energy for the neutral ${\rm Li}$ as a function of the VMC iterations. One can see that the energy decreases  monotonically following the gradient descent optimization and converges for both ansatze in less than 50 iterations. Once the energy has converged within statistical error bars, estimates of the variational energy and cusp parameters are made. 

These results are presented in Table \ref{tablanuclear} alongside  experimental and the most accurate published theoretical results produced by Hylleraas CI (HCI).\cite{BARROIS1997531} One can notice that the difference between the HCI and experimental results appears beyond the fourth significant digit. This is an indication that either the QED corrections from the Breit-Pauli Hamiltonian \cite{PhysRevA.58.3597} or finite mass corrections lie within this order. However, the energies free of corrections are correctly reproduced by both ansatze. Significant differences, beyond statistical errors, appear between the energies predicted by Ansatz A and B for small nuclear charges, but disappear for larger values. The percent energy difference between Ansatz A's predictions and the exact energy lie within roughly 0.01-0.1\% for $Z=3-10$, while that for Ansatz B lies within 0.01-0.05\% over the same range of Z values. 

\subsection{Variationally-Optimized Cusp Parameters}

The cusp conditions (given by Kato's Theorem)
establish that the exact wave function of a system composed of Coulombic charges should reproduce the condition given by Equation (\ref{cusp}) in order to remove divergences in the local energy at the electron-nucleus
coalescence points.\cite{Kato1951FundamentalPO}
Thus, beyond the energy, one observable of interest is the cusp parameter, $C_{Ne}$, which  indicates how capable the electronic wave function is of removing the Coulomb singularity at the position of the nucleus.
The accuracy with which the cusp parameter can be computed is also a good measure of the accuracy of the wave function given its sensitivity to the electronic structure around the nucleus. We find that 98-99\% of the cusp parameter is reproduced by both ansatze, indicating the high quality of the wave functions in the vicinity of the nucleus. Even though the difference between the Ansatze A and B energies is within statistical error bars for $Z=10$, we find that the cusp condition is always better described by Ansatz B.  Satisfying the cusp conditions is a relevant asset in quantum Monte Carlo calculations since it significantly reduces the variance of the local energy during random sampling.\cite{LOOS2019113}
It is also known that fulfilling cusp conditions is relevant to obtain an adequate description of the electron energy distributions in double photoionization.\cite{Drukarev_2007}

\subsection{The Majorana Formula and Parameter Scaling}

With optimized wave function parameters in hand, one may ask how these parameters scale with the nuclear charge. To more deeply appreciate how this scaling differs from previous analytical results, we first compare how our parameters vary with predictions from the Majorana Formula and perturbative treatments to it. 

\begin{widetext}
\begin{figure*}
    \begin{tabular}{cc}
    (a) & (b) \\
    \includegraphics[width=8cm, height=4.8cm,angle=0]{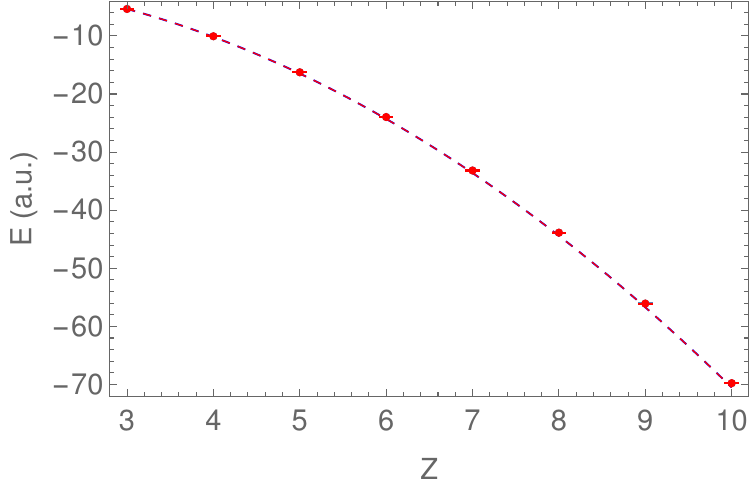} 
   &  \includegraphics[width=8cm, height=5cm,angle=0]{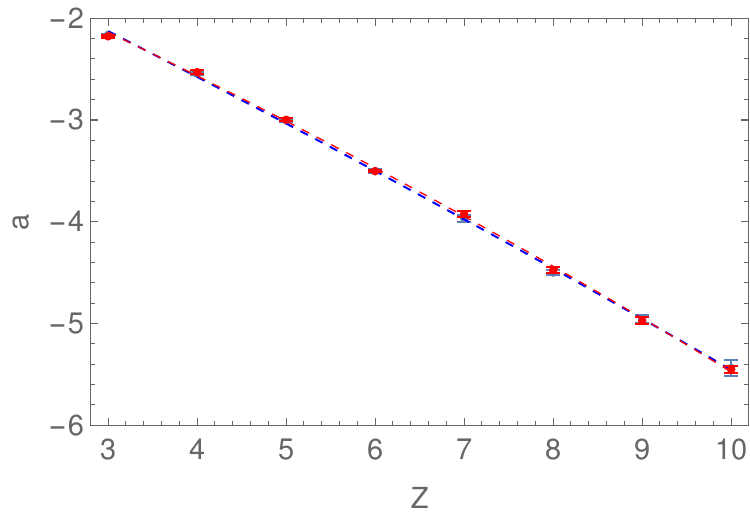} 
   \\ \hline\\
   (c) & (d) \\
  \includegraphics[width=8cm, height=5cm,angle=0]{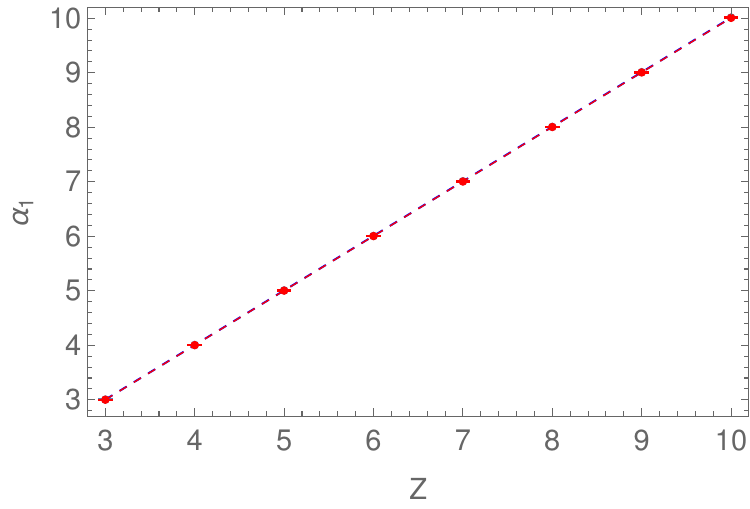} 
   &  \includegraphics[width=8cm, height=4.8cm,angle=0]{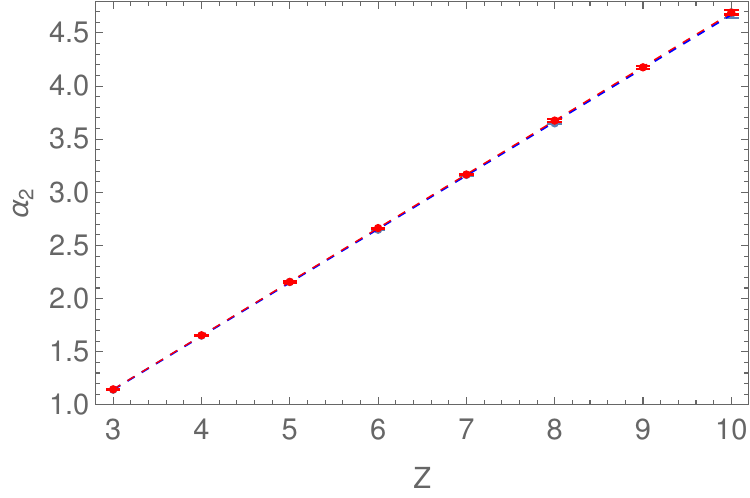} \\
   \hline \\
   (e) & (f) \\
    \includegraphics[width=8cm, height=4.8cm,angle=0]{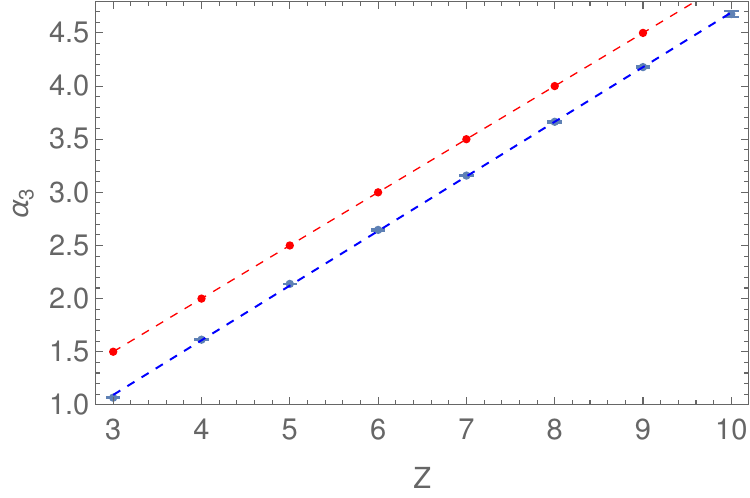} 
   &  \includegraphics[width=8cm, height=4.8cm,angle=0]{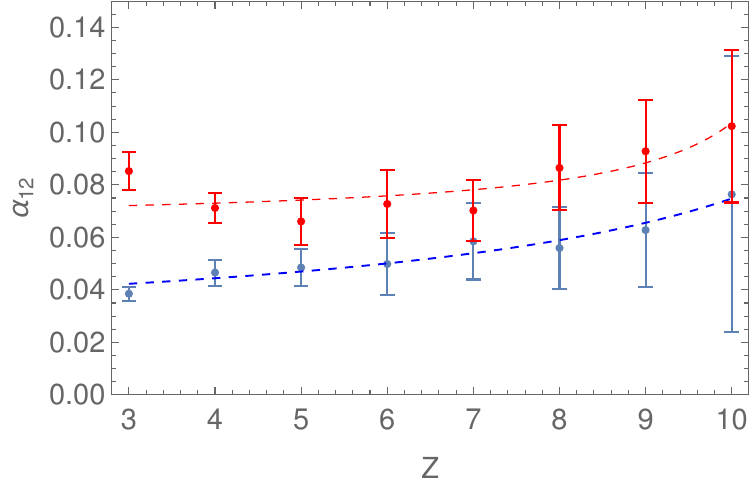} \\
   \hline \\
   (g) & (h) \\
      \includegraphics[width=8cm, height=5cm,angle=0]{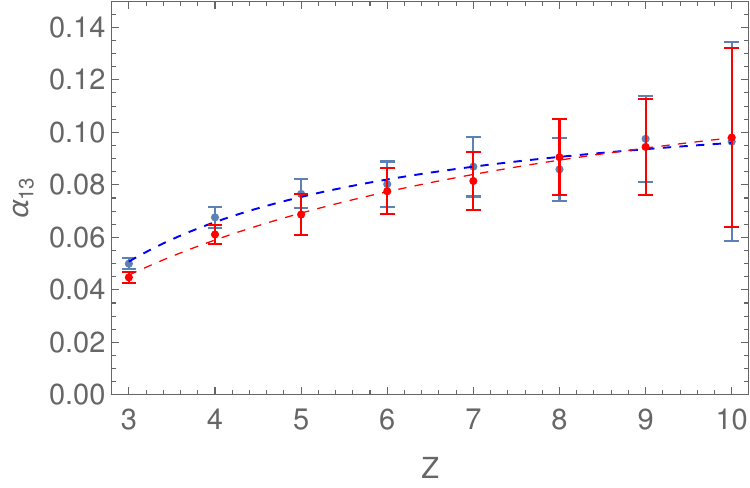} 
   &  \includegraphics[width=8cm, height=5cm,angle=0]{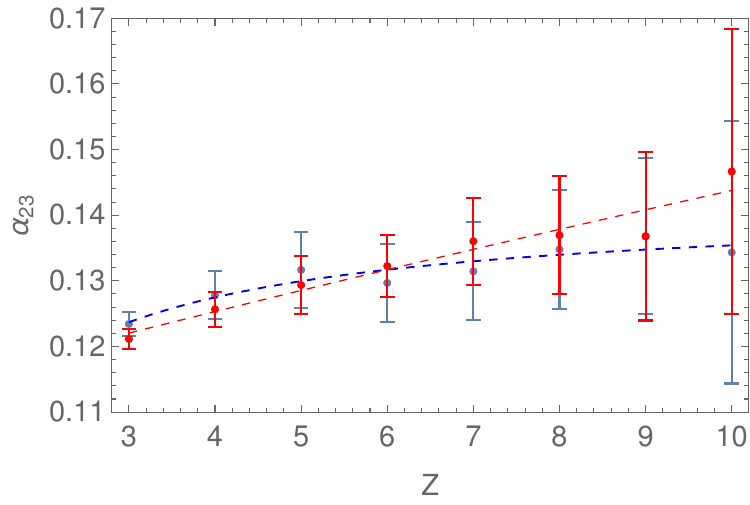} \\
\hline
    \end{tabular}
    \caption{\label{parametros1}
    Energy and variational parameters as a function of the nuclear charge $Z$. The dashed lines correspond to fits to Ansatz A in blue and Ansatz B in red. Error bars are sometimes smaller than the width of the symbol. 
    }
\end{figure*}
\end{widetext}

In his pioneering work that sought a simple, analytical expression for the wave function of helium in its ground state, E. Majorana noticed that the dominant contributions to the ground state energy of He-like atoms comes from a quadratic function of their nuclear charge. This led him to propose the Majorana Formula \cite{Naddeo2012} 
\begin{equation}
\label{Majorana}
E_{M}=-\left(Z-\frac{5}{16}\right)^2\ . 
\end{equation}
This formula can be derived analytically by taking the expectation value of the Hamiltonian with a wave function composed of the product of two 1s orbitals 
\begin{equation}
\label{psimajorana}
\varphi=e^{-\alpha(r_1+r_2)}\,
\end{equation}
with a variational parameter $\alpha$. After variational minimization, the optimal value of $\alpha$ as a function of the nuclear charge, $Z$, is given by 
\begin{equation}
\label{optimalalpha}
\alpha(Z)=Z-\frac{5}{16}\,.
\end{equation}
Note that $\alpha$ varies linearly with the nuclear charge, as is also observed in our plots of our one-body variational parameters in Figure \ref{parametros1}. The Majorana Formula with its optimized $\alpha$ recovers roughly 98\% of the exact energy of helium, an accuracy similar to that provided by Hartree-Fock theory. 

Further corrections to the Majorana Formula can be obtained via the $1/Z$ expansion.\cite{Naddeo2012}
If the wavefunction in Equation (\ref{psimajorana}) is multiplied by an exponential factor 
$e^{\beta r_{12}}\,,$
where $r_{12}$ is the distance between the electrons and $\beta$ another variational parameter, the error of the variational energy decreases to roughly 1\%. Note that this exponential is similar in form to the electron-electron terms present in our Ansatze and these corrections can shed light on how we should expect our electron-electron parameters to scale. Although the integrals needed to estimate the expectation value of the Hamiltonian can be obtained analytically, the optimization of the variational parameters requires estimating the roots of a fifth-degree polynomial which can only be performed numerically.\cite{Abbott_1986} However, one can observe that the following Pad\'e approximant
\begin{equation}
\label{optimalalpha12}
\beta (Z)=\frac{a_0+a_1Z}{b_0+b_1Z}\,,
\end{equation}
correctly describes the optimal values of $\beta$, found in \cite{Abbott_1986}, for nuclear charges between $Z=2-20$. The fitting parameters are $a_0=1.16$, $a_1=-3.11$, $b_0=-2.11$, and $b_1=18.12$. Notice that linear functions are particular cases of Pad\'e functions if $b_0=1$ and $b_1=0$.

In the last column of Table \ref{tablanuclear}, we include the results provided by the Majorana Formula for the lithium 1s2s2p sequence to compare to our variational results. In Figure \ref{parametros1}, we plot (a) the Majorana Formula for the $1s2s2p$ state and (b-l) the optimized wavefunction parameters with error bars as a function of the nuclear charge alongside the fits (whose forms are specified in Table \ref{fits}). It is evident from the plots that parameters $\alpha_{1}$, $\alpha_{2}$, and $\alpha_{3}$ (plots c-e) scale roughly linearly with $Z$, which is in good agreement with Equation (\ref{optimalalpha}). In these cases, the error bars are smaller than can readily be discerned (their exact values can be found in the Supplemental Materials). The other variational parameters are better approximated by Pad\'e functions motivated by Equation (\ref{optimalalpha12}). The fits, including the evaluation of the Majorana Formula, are presented in Table \ref{fits}.

\begin{widetext}
 \begin{center}
\begin{table}[htb!]
\begin{center}
\begin{tabular}{c|ccccccc} \hline
 & \multicolumn{2}{c}{Ansatz A} & \multicolumn{2}{c}{Ansatz B}  \\
$Z$ & $E_v$ & $C_{Ne}$ &  $E_v$ &  $C_{Ne}$ & $E_{HCI}$  & $E_{exp}$ & $E_M$  \\ 
\hline
3  &  -5.3629(1)  & 2.99199(3) &  -5.3650(1)  & 2.99259(3) & -5.3680 & -5.3660   & -5.362 \\
4  & -10.0610(2)  & 3.97481(7) & -10.0631(2)  & 3.97700(2) & -10.0666 & -10.0675 & -10.064\\
5  & -16.2617(4)  & 4.95491(9) & -16.2636(4)  & 4.95798(9) & -16.2676 & -16.2714 & -16.265\\
6  & -23.9636(6)  & 5.9325(1)  & -23.9653(6)  & 5.9375(1)  &  -23.9696 & & -23.967 \\
7  & -33.1659(8)  & 6.9103(2)  & -33.1673(8)  & 6.9119(2)  & -33.1720 & &-33.168 \\
8  & -43.868(1)   & 7.8865(3)  & -43.870(1)   & 7.8912(3)& -43.8748 & &-43.870\\
9  & -56.071(2)   & 8.8616(4)  & -56.072(2)   & 8.8680(4)& -56.0777 & &-56.071\\
10 & -69.773(5)   & 9.8358(5)  &-69.773(5)    & 9.8417(5) &-69.7808 & &  -69.773 \\
 \hline \end{tabular}
\caption{\label{tablanuclear} Variational energies, $E_v$, in Hartrees and cusp parameter, $C_{Ne}$, in atomic units for Li-like ions in the 1$s$2$s$2$p$ quartet state in comparison with the most accurate previous results, $E_{HCI}$ (Hylleraas CI),\cite{BARROIS1997531} and experimental values, $E_{exp}$.\cite{NIST} The last digit in parentheses indicates the statistical error. In the last column, the energies predicted using the Majorana formula, $E_{M}$, are noted.
}
\end{center}
\end{table}
\end{center}
\end{widetext}


\begin{widetext}
 \begin{center}
\begin{table}[htb!]
\begin{center}
\begin{tabular}{c|c} \
Ansatz A & Ansatz B  \\ \hline
$E(Z)=-0.2549+0.5483 Z - \frac{3}{4} Z^2$ & $E(Z)=-0.2580+0.5486 Z - \frac{3}{4} Z^2$\\ \\
 $a(Z)=\frac{-0.8257-0.4153Z}{1-0.0087Z}$ & $a(Z)=\frac{-0.9244-0.3713Z}{1-0.0153Z}$\\ \\
$\alpha_1(Z) =0.0012 + 1.0011 Z$ & $\alpha_1(Z) =-0.0005 + 1.0008 Z$\\ \\
$\alpha_2(Z)=-0.3686 + 0.5040 Z$&$\alpha_2(Z)=-0.3715 + 0.5059Z$\\ \\
$\alpha_3(Z)=-0.4485 +0.5141Z$ & $\alpha_3(Z)=\frac{Z}{2}$ \\ \\
$\alpha_{12}(Z)= \frac{0.0376-0.0013Z}{1-0.0664Z}$ & $\alpha_{12}(Z)= \frac{0.0703-0.0057Z}{1-0.0869Z}$\\ \\ 
$\alpha_{13}(Z)= \frac{-0.3463+0.2314Z}{1+1.9503Z}$ & $\alpha_{13}(Z)= \frac{-0.0388+0.0408Z}{1+0.2769Z}$ \\ \\
$\alpha_{23}(Z)= \frac{0.0512+0.1903Z}{1+1.3434Z}$ & $\alpha_{23}(Z)= \frac{0.1118 + 0.0046Z}{1+0.0099Z}$\\ \\
& $c_3(Z)= \frac{-1.2661+1.7492Z}{10^{-13}+4.6723Z}$\\ \\
& $d_3(Z)= \frac{3.6009 + 8.5600 Z}{1 + 21.0432 Z}$\\ \\
& $c_{12}(Z)= \frac{0.0044 - 0.0078 Z}{1 - 0.0344 Z}$\\ \\
& $d_{12}(Z)=\frac{0.0813 - 0.0065 Z}{1 - 0.0892 Z} $\\ 
\end{tabular}
\end{center}
\caption{\label{fits}
    Fits to the energy (Majorana formula energy given in first row) and to the optimal variational parameters as a function of the nuclear charge. 
    }
\end{table}
\end{center}
\end{widetext}

Also, by analytic continuation of the Pad\'e functions in Table \ref{fits}, we were able to estimate the critical charge \cite{TurbinerCompact3} $Z_c\sim1.1-1.2$ for which Ansatze A and B lose square normalizability, thus predicting the stability of the quartet  state 1$s$2$s$2$p$ for $Z=2$. However,  a closer look at the weakly bound regime is needed in order to extrapolate accurate results for the critical charge in the $Z\rightarrow 0$ limit. 

\section{Conclusions}

In summary, in this manuscript, we optimized ultra-compact, explicitly-correlated wave functions for neutral lithium and its isoelectronic sequence with nuclear charges $Z=3$ to $Z=10$ in the lowest quartet state $1s2s2p$ using the variational Monte Carlo method. We found that VMC yielded highly accurate wave function parameters that resulted in energies
competitive with the accuracy of energies produced by CI ($\sim$95 \% of the correlation energy). We also fit these parameters to linear and Pad\'e functions of the nuclear charge, revealing how they change with charge. 
Uniquely, we have used our VMC approach to study the critical charges of this isoelectronic series, showing that VMC can reproduce 98\% or more of the predicted critical charges. Our results illustrate the power of VMC when it is used along with explicitly-correlated wave functions and indicate that VMC is able to reproduce 99.9\% of the exact energy with only 10 nonlinear variational parameters for the problems studied here. 

We can further improve the accuracy of our wave functions in two ways: by using (i) a linear superposition \cite{Harris2005} of wave functions in the form of Ansatz A with different variational parameters,\footnote{In Ref. \onlinecite{Harris2005}, the authors found that the energy of the {\rm He} ground state converges taking the sum of only four compact, 
explicitly-correlated wave functions.}
or (ii) our optimized wave functions in tandem with other projector QMC methods (typically Diffusion Monte Carlo) that can further improve their structures and minimize their energies. VMC-optimized wave functions are often used as reliable trial wave functions that other, even higher accuracy QMC methods can then refine. 

More work along these lines will need to be done to extend our study to the smaller nuclear charges that posed a challenge in the current work to make our fits more accurate in the weakly-bound regime. 
From there, nuclear charges can be estimated following analytic continuation of the square normalizability constraints. \cite{Turbinercompact2,TurbinerCompact3} The estimate of critical charges in small atomic systems and their behavior close to the threshold is an active area of research. \cite{Burton2021,King2015,MONTGOMERYJR2021}

It is worthwhile to mention that VMC can systematically be extended to more complex systems since its error does not grow exponentially with the system size unlike most high-accuracy methods. However, before adding more electrons to the atomic system, it is also worth examining explicitly-correlated wave functions for doublet excited states with total spin $S=1/2$, where the symmetrization of the total wave function, including the Jastrow factor, is not trivial and therefore spin contamination appears.\cite{Filippi1998}

Furthermore, explicitly-correlated wave functions are useful as starting points to explore the electronic structure of atoms embedded in media \cite{Doma2021} or in magnetic fields. Even if the magnetic field is on the order
typical of those observed in neutron stars, $10^{12}-10^{13}\,$G, and magnetic white dwarfs, $10^8-10^{10}\,$G, the wave function can be approximated by taking the product of the explicitly-correlated wave function in the zero field case and the Landau orbitals for each electron. \cite{TURBINER2006309,TurbinerPRL2013} 

We look forward to the further development and application of these algorithms to tackle these exciting challenges. 

\section{Data Availability}

All codes, scripts, suplemental formulas and data needed to reproduce the results in this manuscript are available online  \hyperlink{https://github.com/djuliannader/VMCLitio}{https://github.com/djuliannader/VMCLitio}.

\section{Acknowledgments}
D.J.N. thanks the kind hospitality of the Brown University Department of Chemistry and A.V. Turbiner and J.C. Lopez Vieyra for insightful discussions. D.J.N. was supported by Fulbright COMEXUS Project NO. P000003405, while B.R. acknowledges support from NSF CTMC CAREER Award 2046744. This research was conducted using computational resources and services at the Center for Computation and Visualization, Brown University.

\section{References}

\bibliography{LiQuartet.bib}

\section{Supplementary Material}

\subsection{Supplemental Equations}

Let us consider the following antisymmetric wave function
 $$\Psi={\mathcal A}\psi=\sum_P{\hat{P}}\psi\,,$$
where ${\mathcal A}$ is the antisymmetrization operator.
For the computation of the local energy $$E_L=\frac{\hat{H}\Psi}{\Psi}\,,$$
one can use the following definition for the kinetic term
\begin{equation}
K_{\psi}\equiv \frac{\hat{T}\psi}{\psi}=-\frac{1}{2\psi} \sum_j^N \nabla^2_j \psi\,,
\end{equation}
where $N$ indicates the number of electrons.
Then the local energy is given by

\begin{equation}
E_L= \frac{1}{\Psi}\sum_P (-1)^P\hat{P}(\psi K_{\psi}) + V\,
\end{equation}
where the sum stands for the permutations of the antisymmetric wave function and $V$ is the potential term.
Note that normalization is not necessary at this point since the local energy is given by the ratio. As for quantities $F_L^p=\frac{\partial_p\Psi}{\Psi}$ and $G_L^p=\frac{\hat{H} \partial_p\Psi}{\psi}$ necessary for the optimization, the derivatives of the wave function with respect to the variational parameters can be conveniently written in terms of the wave function
$$\frac{\partial \psi}{\partial p}=f_p({\bf R})\psi({\bf R}) \equiv \psi_p^\prime ({\bf R})\,.$$

Thus, $F_L^p$ can be calculated simply as
\begin{equation}
\label{Dlocal}
F_L^p=\frac{1}{\Psi}\sum_P (-1)^P\hat{P}(\psi^\prime_p)\,.
\end{equation}

As for $G_L^p$ it can be calculated as
\begin{equation}
 \label{Glocal}
 G_L^p = \frac{\hat{T}}{\Psi}\sum_P (-1)^P\hat{P}(\psi^\prime_p)+\frac{V}{\Psi}\sum_P (-1)^P\hat{P}(\psi^\prime_p)\,.
\end{equation}
The second term is easy to treat since there are no derivatives. However the first term usually leads to complex formulas which may delay the computation time. Therefore it is convenient to reuse some quantities already calculated for local energy $E_L$. To this end we use the following property of the Laplacian operator
$$\nabla^2 (uv)= u\nabla^2 v + v\nabla^2u + 2 \nabla u \cdot \nabla v\,.$$

Thus, the kinetic operator applying on the product $\psi_p^\prime=f_p\psi$ can be calculated as
\begin{eqnarray}
\label{G2}
\hat{T}\psi^\prime_p &=& -\frac{1}{2}\nabla^2 (f_p\psi) =  \psi_p^\prime V_{\psi} + K_{\psi_p} \,,
\end{eqnarray}
defining the quantity
$$K_{\psi_p}=-\frac{1}{2}\psi \nabla^2  f_p - \sum_i^N\nabla_i \psi \cdot \nabla_i f_p\,,$$
where $\nabla^2=\sum_{i=1}^N\nabla^2_i$.
By substituting (\ref{G2}) in (\ref{Glocal}) we finally get
\begin{equation}
\label{Hlocal2}
 G_L^p= \frac{1}{\Psi}\sum_P (-1)^P\hat{P}(\psi^\prime_p(V_\psi+V)) + \frac{1}{\Psi}\sum_P (-1)^P\hat{P}(K_{\psi_p})
\end{equation}

\newpage
\subsection{The algorithm}
\begin{widetext}
\includegraphics[width=1.0\textwidth]{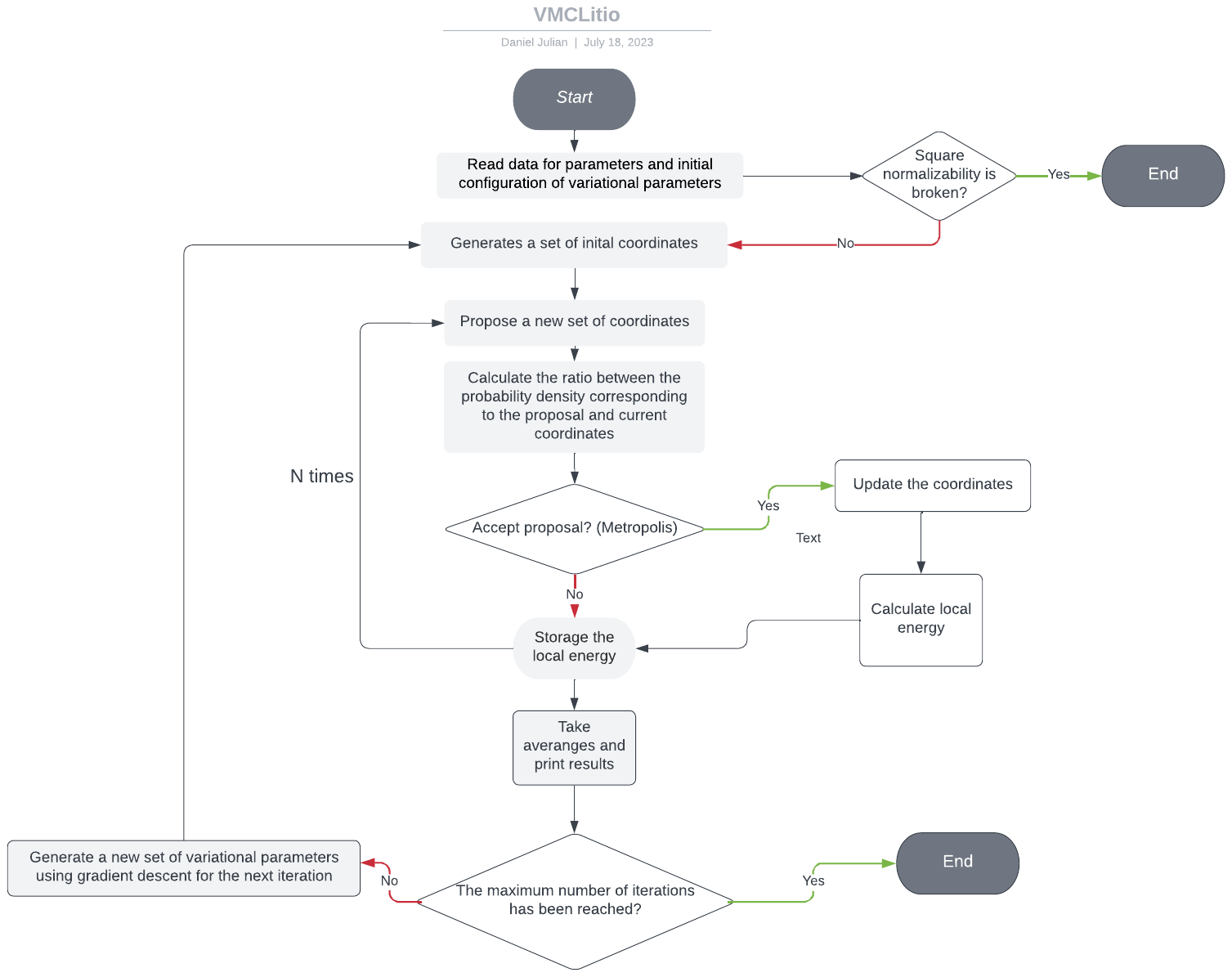}
\end{widetext}
\newpage
\subsection{Using the code}

The code is available on line at {\rm https://github.com/djuliannader/VMCLitio}. We provide a sample script ${\rm Li\_quartetPV3_Z3.dat}$ in the directory examples for optimizing Ansatz A in the case of neutral Lithium.
In the input file one control the following options

\begin{enumerate}
\item Nuclear charge (Z) (double precision)
\item Maximum change in the coordinates (double precision): These variables which control the maximum change possible $\Delta{\bf R}$ in the electronic cylindrical coordinates during the algorithm of Metropolis from one step to the next. They are preferable adjusted in order to get 50 percent of acceptance in the Metropolis algorithm. This can be checked in the acceptance ratio printed on the output file.

\item The steps in the metropolis algorithm (N) (integer)

\item The Subiterations to estimate error (Ms) (integer): each iteration in the Metropolis algorithm is correlated to the previous one. We divide the total iterations N in Ms subgroups in order to estimate the statistical error. Ms must be tuned such that the ratio between Ms and the correlation time is larger than 100 in order to avoid underestimation of the statistical error. The correlation time (AC time) and the ratio between Ms and AC can be checked in the output file.

\item Iterations for the wave function optimization (k) (integer): The code will stop only because the square normalizability is broken or, in the ideal case, because it has reached the desired number of iterations. The convergence of the energy must be checked by following the evolution of the energy as a function of the iterations.

\item Lambda (double precision): corresponds to the parameter of the optimization $\lambda_p$. The code offer the option to set all $\lambda_p$ equally for all parameters.

\item Variational paramaters: In the firs row, the number of variational parameters is set (integer). Then variational parameters are defined in three columns for i) name (string), ii) initial value (double precision) and iii) $\lambda_p$ (double precision). For fixing/releasing parameters just fix $\lambda_p$ as 0.0/1.0. Here, $\lambda_p$ can also be set differently for each parameter in order to accelerate convergence.

We recommend to use the linux command 'grep' to collect data from the output file in order to perform the data analysis.

\end{enumerate}

\newpage 
\subsection{Optimal parameters}

\begin{widetext}
\begin{table}
{\footnotesize
\begin{tabular}{|l|rrrrrrr|}
\hline
$Z$ & $a$ & $\alpha_1$ &  $\alpha_2$  & $\alpha_3$  & $\alpha_{12}$ & $\alpha_{13}$ & $\alpha_{23}$  \\ 
\hline
3 &     -2.162383 $\pm$   0.008481 &  3.004632 $\pm$   0.001098 &  1.142708 $\pm$   0.002206 &  1.067857 $\pm$   0.001800 &   0.038539 $\pm$   0.002715 &  0.049979 $\pm$   0.002159 &  0.123427  $\pm$      0.001874 \\
4    &	   -2.546115 $\pm$   0.013932 &  4.005709 $\pm$   0.001717 &  1.651716 $\pm$   0.003571 &  1.613378 $\pm$   0.002261 &   0.046595 $\pm$   0.004945 &  0.067610 $\pm$   0.004035 &  0.127786  $\pm$	 0.003662 \\
5    &	   -3.004104 $\pm$   0.017922 &  5.007172 $\pm$   0.002360 &  2.156001 $\pm$   0.003990 &  2.138361 $\pm$   0.001412 &   0.048507 $\pm$   0.007091 &  0.076542 $\pm$   0.005523 &  0.131671  $\pm$	 0.005823 \\
6    &	   -3.501891 $\pm$   0.023393 &  6.005965 $\pm$   0.003085 &  2.648104 $\pm$   0.005746 &  2.646436 $\pm$   0.006972 &   0.049847 $\pm$   0.011762 &  0.080252 $\pm$   0.008656 &  0.129676  $\pm$	 0.005997 \\
7    &	   -3.966870 $\pm$   0.035418 &  7.009767 $\pm$   0.003948 &  3.163782 $\pm$   0.007270 &  3.157684 $\pm$   0.007400 &   0.058484 $\pm$   0.014532 &  0.086923 $\pm$   0.011367 &  0.131440  $\pm$	 0.007463 \\
8    &	   -4.500418 $\pm$   0.022157 &  8.006928 $\pm$   0.003679 &  3.649432 $\pm$   0.005644 &  3.663743 $\pm$   0.006972 &   0.055934 $\pm$   0.015517 &  0.085903 $\pm$   0.012035 &  0.134781  $\pm$	 0.009017 \\
9    &	   -4.953553 $\pm$   0.039981 &  9.011290 $\pm$   0.004156 &  4.168285 $\pm$   0.012748 &  4.178627 $\pm$   0.011932 &   0.062774 $\pm$   0.021588 &  0.097529 $\pm$   0.016474 &  0.136818  $\pm$	 0.011938 \\
10   &	   -5.436986 $\pm$   0.077142 & 10.012790 $\pm$   0.005717 &  4.679647 $\pm$   0.039039 &  4.677529 $\pm$   0.026208 &   0.076381 $\pm$   0.052541 &  0.096453 $\pm$   0.037831 &  0.134326  $\pm$	 0.019960 \\
\hline
\end{tabular}
}
\caption{ Optimal variational parameters for Ansatz A }
\label{parametrosA}
\end{table}
\end{widetext}

\begin{widetext}
\begin{table}
{\footnotesize
\begin{tabular}{|l|rrrrrr|}
\hline
$Z$ & $a$ & $\alpha_1$ &  $\alpha_2$  & $\alpha_3$  & $c_{3}$ & $d_{3}$    \\ 
\hline
3      &   -2.178406  $\pm$  0.010804  &   3.001607  $\pm$  0.001829  &  1.146291   $\pm$        0.004588 &   1.500000 $\pm$   0.000000 &   0.281679  $\pm$  0.000357 &   0.456623 $\pm$   0.000295    \\ 
4      &   -2.509496  $\pm$  0.022530  &   4.002861  $\pm$  0.001761  &  1.654847   $\pm$        0.002872 &   2.000000 $\pm$   0.000000 &   0.309589  $\pm$  0.001717 &   0.444298 $\pm$   0.002037    \\ 
5      &   -2.997375  $\pm$  0.016588  &   5.003362  $\pm$  0.002879  &  2.157452   $\pm$	 0.006249 &   2.500000 $\pm$   0.000000 &   0.321483  $\pm$  0.001848 &   0.436247 $\pm$   0.002217    \\
6      &   -3.502820  $\pm$  0.011993  &   6.003769  $\pm$  0.002067  &  2.663565   $\pm$	 0.004611 &   3.000000 $\pm$   0.000000 &   0.332235  $\pm$  0.001475 &   0.432875 $\pm$   0.001709    \\
7      &   -3.924874  $\pm$  0.032010  &   7.004634  $\pm$  0.003534  &  3.167884   $\pm$	 0.007360 &   3.500000 $\pm$   0.000000 &   0.334985  $\pm$  0.002779 &   0.428148 $\pm$   0.003258    \\
8      &   -4.454776  $\pm$  0.031480  &   8.005240  $\pm$  0.003536  &  3.675137   $\pm$	 0.014994 &   4.000000 $\pm$   0.000000 &   0.339463  $\pm$  0.003043 &   0.425345 $\pm$   0.003826    \\
9      &   -4.972842  $\pm$  0.034552  &   9.006199  $\pm$  0.004551  &  4.176393   $\pm$	 0.013668 &   4.500000 $\pm$   0.000000 &   0.342277  $\pm$  0.003908 &   0.423483 $\pm$   0.003341    \\
10     &   -5.453099  $\pm$  0.035319  &  10.007401  $\pm$  0.005884  &  4.693555   $\pm$	 0.020266 &   5.000000 $\pm$   0.000000 &   0.346714  $\pm$  0.006006 &   0.422000 $\pm$   0.006018    \\
\hline
\end{tabular}
}
\caption{ Optimal variational parameters for Ansatz B }
\label{parametrosA}
\end{table}
\end{widetext}

\begin{widetext}
\begin{table}
{\footnotesize
\begin{tabular}{|l|rrrrr|}
\hline
$Z$ &  $\alpha_{12}$ & $c_{12}$ & $d_{12}$ &  $\alpha_{13}$  & $\alpha_{23}$  \\ 
\hline
3  &   0.085239    $\pm$         0.007145 &  -0.016976  $\pm$  0.009876 &   0.092681  $\pm$      0.010557 &  0.044756 $\pm$   0.002210 &   0.121148  $\pm$      0.001482  \\
4  &   0.071189    $\pm$	 0.005877 &  -0.036441  $\pm$  0.007780 &   0.089373  $\pm$      0.014428 &  0.061098 $\pm$   0.003598 &   0.125654  $\pm$	 0.002707  \\
5  &   0.066063    $\pm$	 0.008933 &  -0.038956  $\pm$  0.014612 &   0.080380  $\pm$	  0.012286 &  0.068686 $\pm$   0.007692 &   0.129359  $\pm$	 0.004377  \\
6  &   0.072719    $\pm$	 0.012805 &  -0.061953  $\pm$  0.009531 &   0.088951  $\pm$	  0.008010 &  0.077565 $\pm$   0.008840 &   0.132243  $\pm$	 0.004751  \\     
7  &   0.070235    $\pm$	 0.011524 &  -0.058338  $\pm$  0.012338 &   0.087892  $\pm$	  0.011075 &  0.081469 $\pm$   0.011161 &   0.136028  $\pm$	 0.006636  \\       
8  &   0.086451    $\pm$	 0.016169 &  -0.082975  $\pm$  0.021723 &   0.107738  $\pm$	  0.018015 &  0.090511 $\pm$   0.014532 &   0.136952  $\pm$	 0.008957  \\       
9  &   0.092786    $\pm$	 0.019550 &  -0.091212  $\pm$  0.016389 &   0.119540  $\pm$	  0.013101 &  0.094453 $\pm$   0.018358 &   0.136760  $\pm$	 0.012806  \\  
10 &   0.102344    $\pm$	 0.029069 &  -0.114843  $\pm$  0.014617 &   0.152379  $\pm$	  0.011876 &  0.098013 $\pm$   0.033994 &   0.146654  $\pm$	 0.021733  \\
\hline
\end{tabular}
}
\caption{ Optimal variational parameters for Ansatz B }
\label{parametrosA}
\end{table}
\end{widetext}

\begin{widetext}
\begin{figure*}
    \begin{tabular}{cc}
    (i) & (j) \\
    \includegraphics[width=8cm, height=5cm,angle=0]{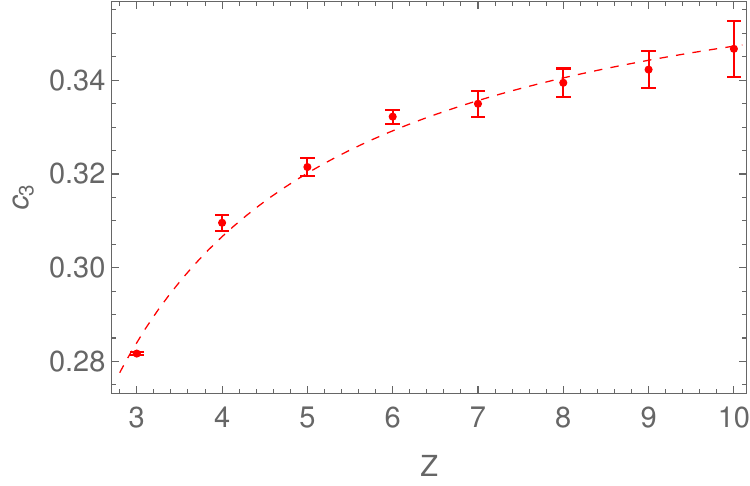} 
   &  \includegraphics[width=8cm, height=5cm,angle=0]{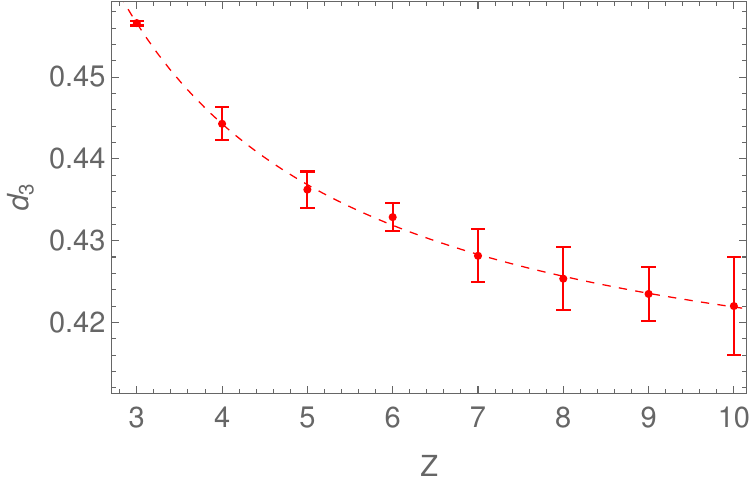} \\ \hline \\
   (k) & (l) \\
  \includegraphics[width=8cm, height=5cm,angle=0]{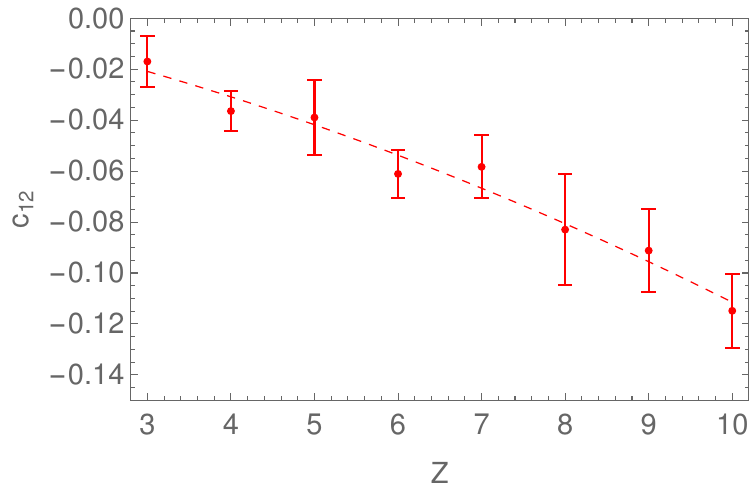} 
   &  \includegraphics[width=8cm, height=5cm,angle=0]{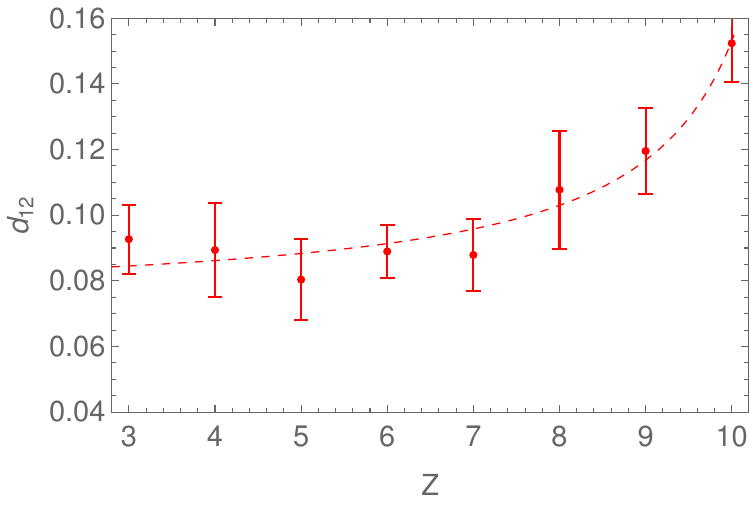} \\
\hline
    \end{tabular}
    \caption{\label{parametros2}
    Continuation of Figure \ref{parametros1}.
    }
\end{figure*}
\end{widetext}

\end{document}